\documentclass[aps,apl,twocolumn,superscriptaddress]{revtex4-1}
\usepackage{amssymb}
\usepackage{amssymb}
\usepackage{amssymb}
\usepackage{graphicx}
\usepackage{graphics}
\usepackage{amsmath}
\usepackage{placeins}
\usepackage{hyperref}
\usepackage[dvipsnames]{xcolor}
\usepackage{wasysym}

\usepackage{soul}
\begin{document}

\title{Demonstration of Planar Ultrananocrystalline Diamond Field Emission Source Operating in SRF Injector at 2 Kelvin}
	\author{\firstname{Sergey V.} \surname{Baryshev}}
	\email{serbar@msu.edu}
	\affiliation{Department of Electrical and Computer Engineering, Michigan State University, East Lansing, MI 48824, USA}
	\author{\firstname{Erdong} \surname{Wang}}
	\affiliation{Brookhaven National Laboratory, Upton, NY 11973, USA}
	\author{\firstname{Chunguang} \surname{Jing}}
	\affiliation{Euclid TechLabs, Bolingbrook, IL 60440, USA}
	\author{\firstname{Vadim} \surname{Jabotinski}}
	\email{Currently at Leidos}
	\affiliation{Euclid TechLabs, Bolingbrook, IL 60440, USA}
	\author{\firstname{Sergey} \surname{Antipov}}
	\affiliation{Euclid TechLabs, Bolingbrook, IL 60440, USA}
	\author{\firstname{Alexei D.} \surname{Kanareykin}}
	\affiliation{Euclid TechLabs, Bolingbrook, IL 60440, USA}
	\author{\firstname{Sergey} \surname{Belomestnykh}}
	\email{Currently at Fermilab}
	\affiliation{Brookhaven National Laboratory, Upton, NY 11973, USA}
	\author{\firstname{Ilan} \surname{Ben-Zvi}}
	\affiliation{Brookhaven National Laboratory, Upton, NY 11973, USA}
	\author{\firstname{Lizhi} \surname{Chen}}
	\email{Currently at Sarah Lawrence College}
	\affiliation{Brookhaven National Laboratory, Upton, NY 11973, USA}
	\author{\firstname{Qiong} \surname{Wu}}
	\affiliation{Brookhaven National Laboratory, Upton, NY 11973, USA}
	\author{\firstname{Hao}
	\surname{Li}}
	\affiliation{Center for Nanoscale Materials, Argonne National Laboratory, Lemont, IL 60439, USA}
	\author{\firstname{Anirudha V.}
	\surname{Sumant}}
	\affiliation{Center for Nanoscale Materials, Argonne National Laboratory, Lemont, IL 60439, USA}

\begin{abstract}
Reported here is the first demonstration of electron beam generation in an SRF TESLA 1.3 GHz gun equipped with field emission cathode when operated at 2 Kelvin. The cathode is submicron film of nitrogen-incorporated ultrananocrystalline diamond [(N)UNCD] deposited atop a Nb RRR300 cathode plug. The output current was measured to increase exponentially as a function of the cavity gradient. Our results demonstrate a feasible path toward simplified fully cryogenic SRF injector technology. One important finding is that the electron emitter made of (N)UNCD, a material long been known as a highly efficient field emission material, demonstrated a record low turn-on gradient of 0.6 MV/m. A hypothesis explaining this behavior is proposed.
\end{abstract}

\maketitle

Superconducting radiofrequency (SRF) technology is becoming the technology of choice for many scientific projects such as ion-electron colliders (RHIC/eRHIC), nuclear accelerators and cyclotrons (FRIB and NSCL), free-electron lasers (LCLS-II), and proton accelerators (PIP-II). Large investments in SRF technology pay off in that the overall mass production quality of 100’s MHz to GHz niobium cavity technology has significantly improved. As a result, SRF technology becomes an attractive option for industry. Industrial applications of SRF accelerators include (but not limited to) portable X- to $\gamma$- ray sources for cargo scanning, sterilization, radioisotope and radiopharmaceutical production. The main drivers and benefits behind industrial SRF systems are energy efficiency (about 50\% wall plug efficiency) and high production rate (high duty cycle or CW operation). The use of traditional thermionic cathodes or photocathodes combinations may compromise these two benefits, or, in case of photocathode injector, additionally complicate and price up the operation and maintenance. A natural way to simplify (and reduce operation cost) and integrate SRF architecture with the electron source would be to place the source directly into the SRF cavity. Since photo- and thermionic- cathodes have a number of disadvantages, a cold and dark cathode compatible with operating temperatures of a few K is highly desirable. The latter implies that field emission electron source should be attempted. Finding an electron field emission source able to sustain a high repetition rate or CW operation and to provide high current is therefore the most critical challenge.

To address this challenge, in this work we explore highly conductive nitrogen-incorporated ultrananocrystalline diamond [(N)UNCD] as a field emission cathode material that could complement and simplify fully cryogenic SRF injector designs. (N)UNCD has been proven to be a highly emissive material when tested in normal conducting high gradient L- and X-band injectors \cite{JiaqiIEEE2018,SergeyAPL2014}. Electron emission from the (N)UNCD planar surface with excellent emittance, energy spread, and stability were confirmed. rf pulse currents of $\sim$1-100 mA were achieved \cite{SergeyAPL2014, JiaqiIEEE2018, Shao2019}. At high rep-rate/CW operation, such pulse current range serves as an average current estimate for SRF applications, given that field emission is a quantum tunneling effect (meaning that similar currents are expected between 2 K and room temperature). In the present paper, we describe a methodology of integration of a (N)UNCD field emission cathode into a half-cell L-band 1.3 GHz SRF gun, as well as illustrate basic performance characteristics obtained at 2 K.

\begin{figure*}
\includegraphics[height=4.5cm]{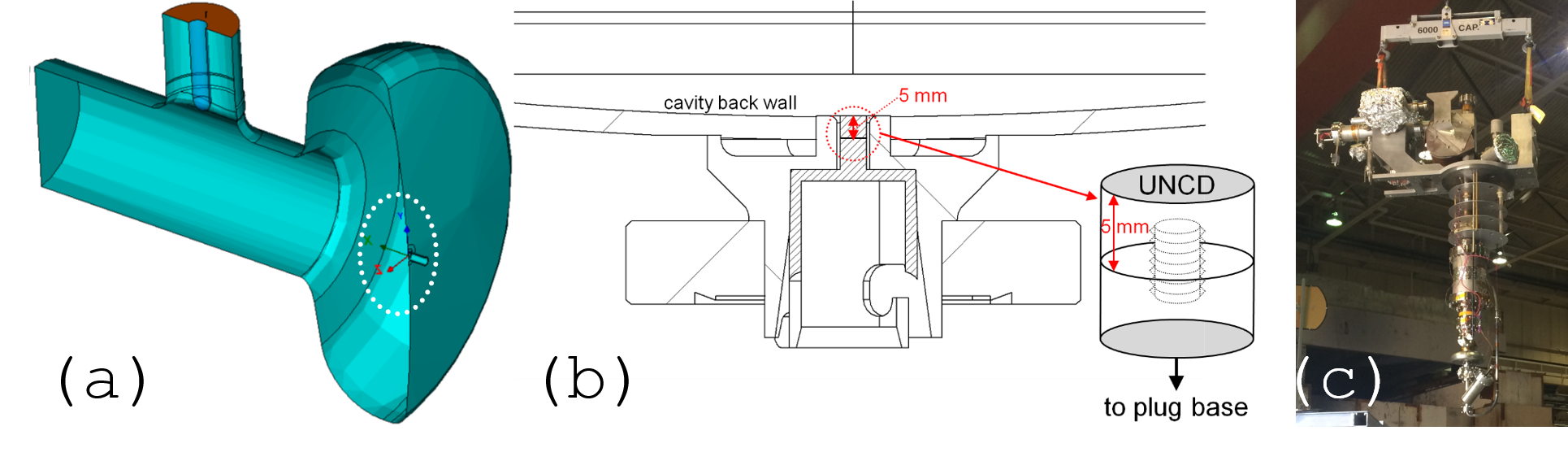}
\caption{\label{f1}(a) Cut-away view of 3D model of the 1.3 GHz superconducting (SRF) cavity with field emission cathode installed in its back wall (dashed circle); (b) Plug insertion zone on the back wall of the SRF cavity (the screw joining between the plug base and the emitter cartridge is shown to the right); (c) The entire beamline being installed into the cryomodule.}
\end{figure*}

The 3D model of the half-cell TESLA injector used in this study is shown in Fig.\ref{f1}a. It is designed for photocathode studies \cite{ErdongERL}. For this reason, the cathode plug has a unique design: it consists of a bulky base (approx. 30 mm wide) with a mechanical lock notches and pins, and a narrow cylindrical tip 4 mm in diameter and 8 mm in length. The surface of the tip should be covered with an emitter material. The plug is locked such that the tip is placed flush with respect to the inner back-wall surface of the cavity. Zoom-in cathode installation zone in the cavity is shown in Fig.\ref{f1}b.

The niobium plug is too large to be placed into a chemical vapor deposition (CVD) reactor for tip coating with (N)UNCD. Diamond CVD reactors are typically designed for thin film deposition onto planar microelectronics substrates. First of all, the substrate has to be uniformly (no large gradients) heated to a required growth temperature (800-850 $^\circ$C in our case) -- this is troublesome with such a large hollow metal piece. In the case of the microwave-assisted CVD system we employ, high/large conductive substrates can be accommodated, thanks to an automatic impedance matching network. However, the microwave field distribution over the growth area needs to be uniform for uniform growth -- this is troublesome with such a high aspect ratio cathode. Therefore, the top section of the cathode plug had to be detachable. Similarly to previous L-band and X-band cathode designs \cite{JiaqiIEEE2018,SergeyAPL2014}, the top 5 mm of the 8-mm tip was cut: this could allow for quick emitter cartridge sample exchanges for testing. Three options were considered for re-attaching the (N)UNCD coated top section: electron beam welding, indium/lead/tin soldering, mechanical joining. Electron welding was rejected as it could induce temperature of the tip to up to $>$1,000 $^\circ$C which would convert (N)UNCD into nano-graphite. Soldering with In, Pb and Sn was considered because these metals are superconductors and the melting point of their combinations can be lower than 200 $^\circ$C. This option was rejected as such joining could be too fragile and could therefore introduce particulates or contamination to the Nb cavity.

\begin{figure}
\includegraphics[height=8cm]{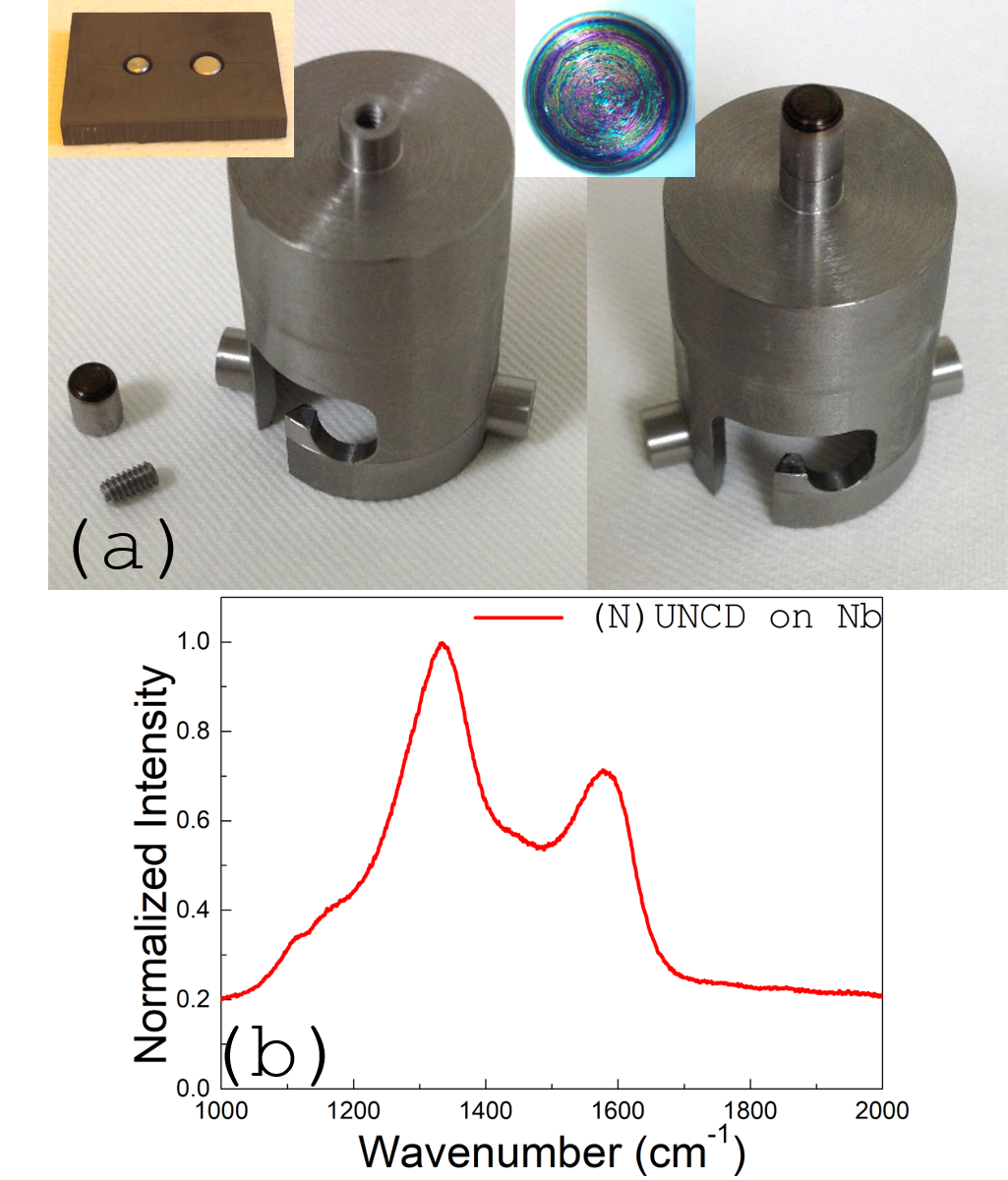}
\caption{\label{f2}(a)The Nb cathode plug: in disassembled and assembled form. The insets are the pyrolytic graphite mask hosting the cathode tip cartridges and the top view of the emitting (N)UNCD film coating atop the Nb substrate; (b) Raman spectrum confirming (N)UNCD structure.}
\end{figure}

The mechanical joining using a 2-56 Nb set screw was chosen in order to provide robust electrical and mechanical contact between the emitter cartridge and the plug base (see Figs.\ref{f1}b and \ref{f2}a). The cut top 5 mm of the 8-mm tip and the counter part were drilled and tapped. A set of emitter cartridges and set screws were machined out of Nb RRR300. All parts went through standard ultrasonic bath cleaning followed by the BCP (buffered chemical polish) process. A slower etching BCP recipe of HF(48\%):HNO$_3$(70\%):H$_3$PO$_4$(85\%) = 1:1:2 was chosen. The parts were put into the BCP bath for 1 min at room temperature and rinsed with deionized water. Approximately 5 $\mu$m of material was removed. Next, a specialty mask (inset in Fig.\ref{f2}a) of thickness of 5 mm (matching the cathode cartridge height) was designed and machined out of pyrolytic graphite. Graphite mask surface was flush with the cartridge surface helping avoid enhancement of the 915 MHz microwave field (plasma discharge in the CVD reactor) at the plug edges and therefore achieve laterally uniform growth of the (N)UNCD film over the substrate. Growth time was 40 minutes which corresponded to $\sim$200 nm. Raman spectroscopy confirmed that the film had canonical (N)UNCD structure (Fig.\ref{f2}b). The plug assembly in final pre-installation form is shown in Fig.\ref{f2}b.

\begin{figure*}
\includegraphics[height=5cm]{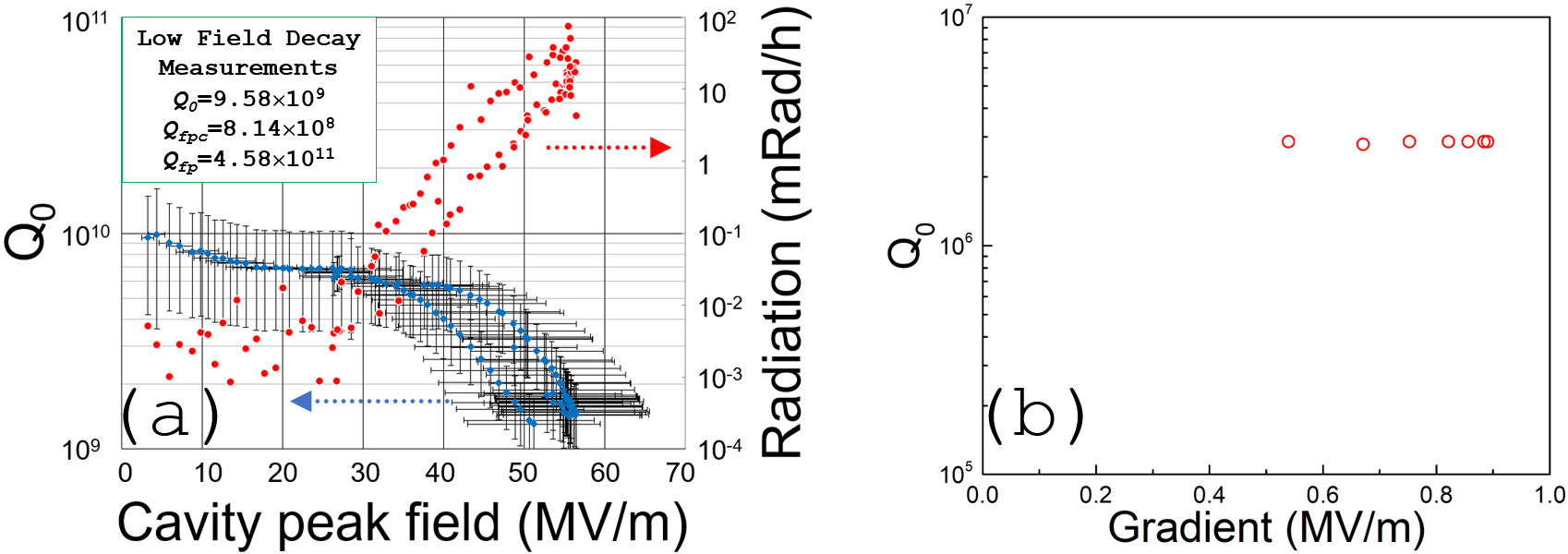}
\caption{\label{f3}. (a) $Q_0$ of the gun as a function of the cavity peak field with uncoated Nb cathode plug versus (b) $Q_0$ as the function of cathode gradient with (N)UNCD-coated Nb cathode plug.}
\end{figure*}

The entire gun is vertically immersed into the liquid helium bath (Fig.\ref{f1}c). The beamline has three ion pumps and two NEG pumps to maintain $10^{-11}$ Torr in the gun. The cathode was installed and locked with 10 lbs force applied for maintaining sufficient thermal conduction between cathode plug and gun. Table 1 shows the gun intrinsic rf parameters, i.e. without (N)UNCD. With the standard, uncoated, Nb plug, $Q_0$ of the gun is $7\times 10^9$ at 2 K yielding maximum field in the cavity of 56 MV/m as shown in Fig.\ref{f3}a. Dark current emission threshold is 30 MV/m. The input power coupler has $Q_{fpc}$ of $8\times 10^8$ while the pickup coupler has $Q_{fp}$ of $4.5\times 10^{11}$. Both couplers are fixed to match the nano-ampere beam current. Initially, we used the same coupling settings to test the (N)UNCD cathode.

\begin{table}[!]
\caption{Intrinsic parameters of the 1.3 GHz SRF gun (i.e. w/o (N)UNCD emitter coating)}
\begin{ruledtabular}
\begin{tabular}{lcc} 

 rf frequency & 1.3 GHz \\
 cavity active length & 0.6 cell \\
 energy gain & 0.6 MeV \\
 maximum field at the cathode & 15 MV/m \\
 cavity operating temperature & 2K \\
 cavity $Q_0$ at 2 K & $7\times 10^9$ \\
 bunch repetition frequency & 1.3 GHz
 
\end{tabular} 
\end{ruledtabular}
\label{tb:1}
\end{table}

To obtain current measurement, the gun was electrically floated. Both cathode insertion port and beam exit port have ceramic insulation. An individual dc blocker was installed on each coaxial coupler cable. One wire was attached to the gun flange and connected to an oscilloscope measuring current from the gun. We used a self-exciting loop to control rf operation: it means the beamline diagnostics are synchronized with the gun operating frequency. All cable calibrations were done at room temperature. The gun was first cooled down to 4 K and then cooled down to 2 K using liquid helium pumping. Decay-time measurement was employed to determine $Q_0$ of the gun with (N)UNCD: it changed to $1.74\times 10^6$ at 4 K and $2.3\times 10^6$ at 2 K. The actually measured gun quality factor versus cathode gradient is shown in Fig.\ref{f3}b. Reduced $Q_0$ resulted in deteriorated coupling between the gun and rf source: 99\% of rf power was reflected at the main coupler. The maximal available power of 100 W from the solid-state amplifier and the weak coupling limited the gun gradient.

\begin{figure}
\includegraphics[height=5.5cm]{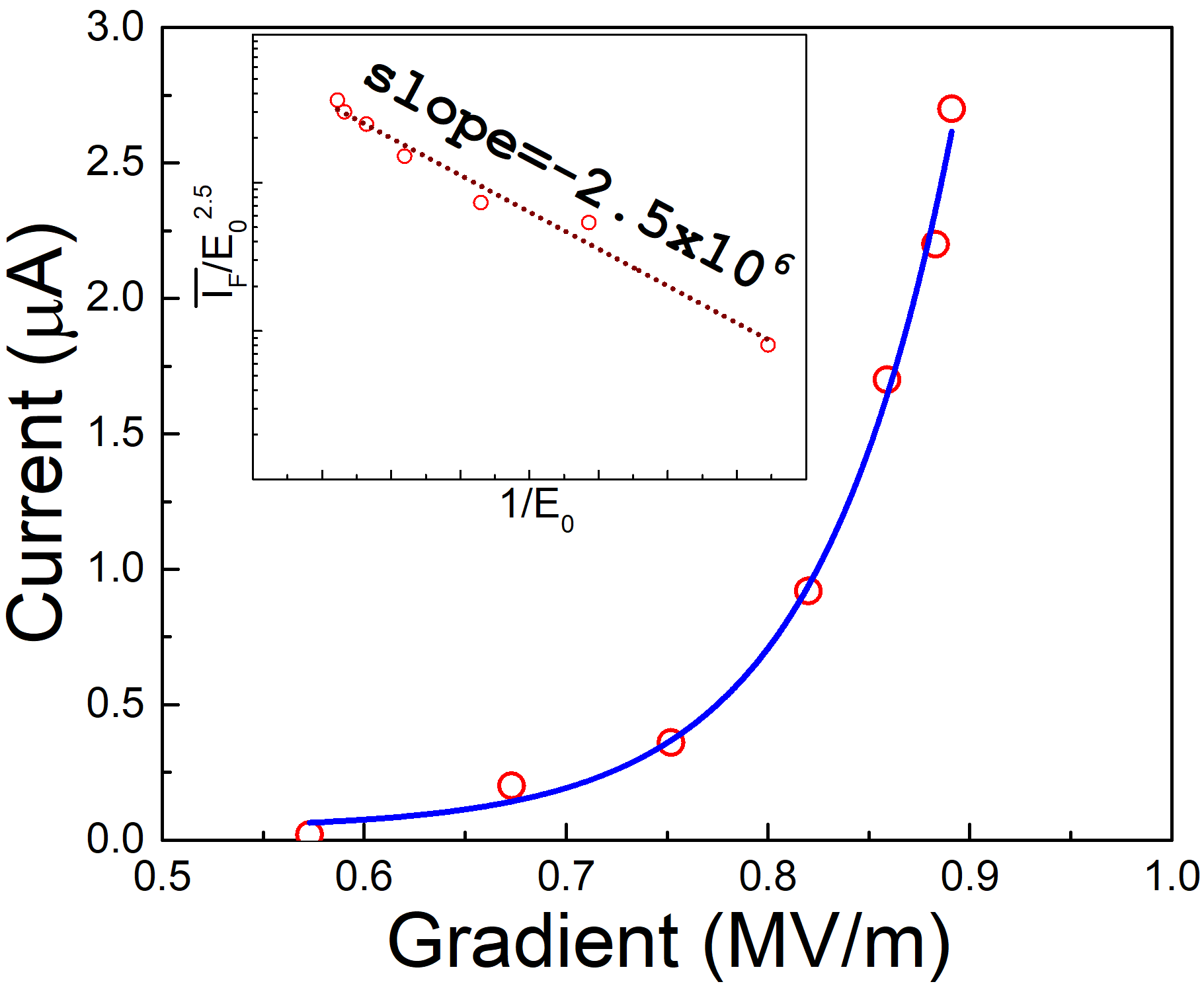}
\caption{\label{f4}The time-averaged output current as a function of the cathode gradient measured at 2 K. The inset shows the FN-like plot used for slope $s$ extraction.}
\end{figure}

No output beam signal was detected at input power of 90 W from the solid-state amplifier. By maximally increasing the coupler length, the $Q_{fpc}$ was reduced to $2\times 10^7$ which was still far (order of magnitude) from optimal matching and caused significant rf loss. Time-domain reflectometry indicated two reflections in the input power cables: one was at the dc blocker and another one was in the input coupler which dominated the power reflection.  However, this optimized setting was enough to obtain a measurable field emission current (Fig.\ref{f4}).

In the final experiments, it was impossible to transport the beam to the Faraday cup located approx. 1.5 m away, at the end of the beamline: the resulting beam energy was 0.03 MeV, i.e. much lower compared to the anticipated energy of 1.4 MeV. Therefore, the electron beam current leaving the gun was directly measured on the oscilloscope. Steadily increasing the input power, the current as a function of the cathode gradient was established. The dependence is plotted in Fig.\ref{f4} -- the behavior is exponential and indicates that the emission current retains Fowler-Nordheim (FN) like behavior. As shown in Fig.\ref{f3}a, intrinsically the gun with the standard uncoated Nb cathode starts emitting dark current at gradients over 30 MV/m. This corroborates that the introduction of the (N)UNCD material drastically changes the gun emission behavior, and the (N)UNCD emitter material is likely responsible for the emitted current. The most remarkable and surprising result was that the emission threshold gradient was as low as 0.57 MV/m. Even though a turn-on field as low as 0.5 MV/m was reported once for nitrogen doped microcrystalline diamond in dc regime \cite{Nature}, in our previous dc and rf studies \cite{ChubenkoACS2017,Shao2019} lowest turn-on dc field and rf gradient were 2.5 and 8 MV/m, respectively, for the (N)UNCD emitters synthesized under same conditions. In case of the latest rf measurements \cite{Shao2019}, the rf system was limited in terms of lowest power (yielding gradient on the cathode), and pulse length and duty cycle (yielding total emitted charge per second) available from the L-band klystron. Therefore, the record low emission threshold obtained in these rf measurements merits further discussion.

Since the detecting scope bandwidth was 200 MHz, and the signal cable was not matched between the gun and outside environments, the output current signal was averaged as
\begin{equation}\label{e1}
\overline{I_{F}}=\frac{\int_{0}^{T}{I_F(t)dt}}{T},
\end{equation}
where $T$ is the rf period and $I_F(t)$ is the field emission current at time $t$. The classical FN law describes the response to the dc field. A modified version of the FN law can be proposed to describe the averaged rf field emission current as a function of the cathode gradient. If $\omega$ is the operating rf frequency and the cathode gradient has time dependence as $E=E_0$sin($\omega t$) with $E_0$ being the amplitude/peak gradient of the macroscopic surface field on the cathode, averaging over the period yields \cite{JuwenSLAC1997}
\begin{equation}\label{Eqn_FN_ave}
	\begin{aligned}
	\overline{I_{F}}=&\frac{5.7\times 10^{-12}\cdot10^{4.52\phi ^{-0.5}}\cdot A_{e}\cdot(\beta\cdot E_{0})^{2.5}}{\phi ^{1.75}}\\
	&\cdot \exp[-\frac{6.53\times 10^{9}\cdot\phi ^{1.5}}{\beta\cdot E_{0}}],
	\end{aligned}
	\end{equation}
where $\beta$, $\phi$ and $A_e$ are the field enhancement factor, work function and emission area respectively. If the common logarithm of $\overline{I_{F}}/E_0^{2.5}$ versus $1/E_0$ is plotted a relation between $\beta$ and $\phi$ can be obtained as
	\begin{equation}\label{Eqn_beta}
	\frac{\phi ^{1.5}}{\beta}=-\frac{s}{2.84\times 10^{9}},
	\end{equation}
	where $s$ is the slope of the linear dependence (see the inset in Fig.\ref{f4}). The slope $s$ is extracted from the modified FN plot but the field enhancement factor and work function are correlated. (N)UNCD emitters are planar (intrinsic UNCD roughness is $\sim$10 nm) with surface roughness limited by the quality of mechanical polishing of the metal substrate (typical substrate $\beta$'s$\sim$1 \cite{Shao2019}), and therefore in this emitter system the field enhancement factor does not have standard geometrical interpretation. It is formally extracted from the FN plot when the work function is known from an independent measurement using a Kelvin probe or photoelectron spectroscopy method. Previously we measured the work function of films grown identically to be within 3.6 \cite{PerezAPL2014} to 4 eV \cite{Chen2019} at room temperature. When the environment temperature is reduced by two orders of magnitude, from 300 to 2 K, Nb tip is expected to compress (it has fairly large thermal expansion coefficient), thus exerting stress on the (N)UNCD emitter coating that may cause compression strain. It was computationally and experimentally shown that in materials undergoing compression strain work function can lower significantly, 0.5 to 1 eV \cite{wf1,wf2}. Therefore, we are testing this hypothesis through the following approach.
	
	In our recent conditioning rf experiments, it was established that the field enhancement factor is not conserved in the system as the field emitter is progressively exposed to higher gradient, it gradually reduces \cite{Shao2019}. The product of the field enhancement factor and the maximum surface gradient at which this field enhancement factor was measured is conserved -- this product of $\beta\cdot E_0$ is $\approx$5-7.5 GV/m. This metrics is similar to 10 GV/m obtained for copper in dc experiments as a product of $\beta$ and the breakdown field $E_b$ \cite{CERN}. The established phenomenological relation between $\beta$ and $E_0$ can be used to independently extract $\beta$ of our sample (conditioned to 0.9 MV/m) to be equal to $(5.5\div 8.3)\times 10^3$. Then from Eq.(3), it can be found $\phi$ ranges from 2.9 to 3.8 eV. The extracted work function values are not unusual and agree well with previously measured range 3.6-4.0 eV,  keeping in mind possible compression strain effects. Unlike low duty cycle pulsed rf operation of (N)UNCD in L-band copper injector, operation in CW makes it more sensitive to lower emitted charge and results in earlier current detection (lower turn-on gradient).
	
	In conclusion, a feasible approach of operating an SRF injector with field emission electron source was demonstrated. Before this or similar approach can become practical, a number of issues need to be solved. For example, strain-mediated effect on work function is a potentially interesting effect to study to fully characterize field emission cathode operated at cryogenic temperatures. More practically, it was shown that $Q_0$ of the gun with (N)UNCD was three orders of magnitude lower than that with pure Nb plug. This could be caused by dielectric or ohmic loss that would increase the local temperature and quench the quality factor by forcing the Nb tip into the normal conducting state even at very low gradient. Fig.\ref{f2}a shows that (N)UNCD film coating extends outwards to the sides. A previous study showed that the heat dissipation at the cathode edges dominates the rf power loss \cite{ErdongCPhys}: reducing the diamond coated area could help avoid edge-dominated power dissipation. Additionally, the Nb tip was separated into two pieces for synthesis and rejoined with a Nb set screw. This design could reduce the thermal conductivity and additionally contribute to forcing the tip into the normal conducting state.
	
\textbf{Acknowledgment.} We would like to thank Allan Rowe and Anthony Crawford (FNAL) for their help with BCP processing. This work was supported by Nuclear Physics DOE SBIR Grant No. DE-SC-0013145. The use of the BNL SRF test facility was made possible through High Energy Physics DOE Stewardship Grant “Ultra-Nanocrystalline Diamond Cathode Testing in an SRF gun”. (N)UNCD deposition and its characterization was performed at the Center for Nanoscale Materials, a DOE Office of Science User Facility, and supported by DOE, Office of Science, under Contract No. DE-AC02-06CH11357. Part of this work was funded by DOE-TCF grant ED2701000-05450-1009209.

\bibliography{srf}

\end{document}